\documentclass[12pt,aps,noshowpacs]{revtex4}
\usepackage{amssymb}
\usepackage{amsmath}
\usepackage{amsfonts}
\usepackage{graphicx}

\input{tcilatex}
\begin{document}

\title{Gravitational Radiation Observations }
\author{E.N. Glass }
\affiliation{Department of Physics, University of Michigan, Ann Arbor, MI 48109}
\date{September \ 25, 2016}
\maketitle

\subsection{Electromagnetic and Gravitational Radiation}

The notion of gravitational radiation begins with electromagnetic radiation.
In 1887 Heinrich Hertz, working in one room, generated and received
electromagnetic radiation. Maxwell's equations describe the electromagnetic
field. The quanta of electromagnetic radiation are spin 1 photons. They are
fundamental to atomic physics and quantum electrodynamics.

Since the field concept underlies electromagnetic radiation, it was assumed
that a similar concept gave rise to gravitational radiation. The
gravitational field equations are Einstein's equations of General
Relativity. The weakness of the gravitational interaction is epitomized by
Newton's constant "G" which is known to only 3 significant figures. By
analogy with electrodynamics, one refers to gravitational quanta as spin 2
gravitons. But the gravitational field is purely classical, not quantum.

\subsection{Early measurements}

Gravitational radiation arises from distant astrophysical events. In 1960
Joseph Weber, a physicist at the University of Maryland, began an experiment
to detect gravitational waves. Weber and his students used "Weber bars". The
"bars" consisted of multiple aluminium cylinders, 2 meters in length and 1
meter in diameter, antennae for detecting gravitational waves. These 1.2 ton
aluminium cylinders vibrated at a resonance frequency of 1660 hertz and were
designed to be set in motion by the gravitational waves predicted by Weber.
Because these waves were supposed to be weak, the cylinders had to be
massive and the piezoelectric sensors had to be very sensitive, capable of
detecting a change in the cylinders' length of about 10$^{-16}$ meters.
Weber claimed to have detected gravitational waves. Other groups began to
build their own devices. Over a period lasting several years, the other
groups all failed to replicate Weber's results with their instruments. In
the 1970s, the results of these gravitational wave experiments were largely
discredited, although Weber continued to argue that he had detected
gravitational waves. In order to test Weber's results, IBM physicist Richard
Garwin built a detector that was similar to Weber's. In six months, it
detected only one pulse, which was most likely noise.

\subsection{Recent measurements}

Gravitational waves have been detected by LIGO \cite{Abb1,Abb2} (Laser
Interferometer Gravitational Wave Observatory) with interferometers at
Hanford, Washington and Livingston, Lousiana (3,700 km apart ). LIGO's
interferometers are the largest ever built. With arms 4 km long, they are
360 times larger than the interferometer used in the Michelson-Morley
experiment.

Resonant mass gravitational wave detectors like Weber bars and
interferometric detectors such as LIGO look for the same effect: the
stretching and compressing of space caused by a gravitational wave. Bars are
far less expensive than detectors like LIGO but are only sensitive to narrow
ranges of gravitational wave frequencies. Bars are also only sensitive to
small portions of the sky at once where detectors like LIGO are sensitive to
most of the sky at once (including the sky on the other side of the planet).
Because of how each of these detectors look for gravitational waves,
resonant-mass detectors are likely to only be sensitive to the strongest
gravitational waves.

On September 14, 2015, both LIGO "antennas" made the first direct
measurement of gravitational waves. The event was detected in coincidence by
the two antennas. Physicists have concluded that the detected gravitational
waves were produced during the final fraction of a second of the merger of
two black holes to produce a single, more massive spinning black hole. This
collision of two black holes had been predicted but never observed. Based on
the observed signals, LIGO scientists estimated that the black holes for
this event were about 29 and 36 solar masses, and the event took place about
1.3 billion years ago. Nearly 3 solar masses were converted into
gravitational waves in a fraction of a second with a peak power output
roughly 50 times that of the whole visible universe

The gravitational wave arrived at the two detectors at almost the same time,
indicating that the source was located somewhere in a ring of sky about
midway between the two detectors. Knowing the LIGO detector sensitivity
pattern, it was a bit more likely overhead or underfoot instead of to the
West or the East. With only two detectors one can't narrow it down much more
than that. This differs from LIGO's first detected signal (September 14,
2015), which came from the 'southeast', hitting Louisiana's detector before
Washington's.

The two merging black holes were less massive (14 and 8 solar masses) than
those observed in the first detection (36 and 29 solar masses). While this
made the signal weaker than the original one, when these smaller black holes
merged, their signal shifted into higher frequencies bringing it into LIGO's
sensitivity band earlier in the merger than was observed in the September
event. This allowed more orbits to be observed than the first
detection--some 27 orbits over about one second (compared with just two
tenths of a second of observation in the first detection). Combined, these
two factors (smaller masses and more observed orbits) were the keys to
enabling LIGO to detect a weaker signal. They also allowed one to make more
precise comparisons with General Relativity.

\subsection{Future work}

Researchers are planning and building the next generation of larger and more
isolated detectors deep beneath the ground where hundreds of meters of
overlying rock shield against most human noises and seismic stresses. In the
Kamioka mine in Japan, the Kamioka Gravitational Wave Detector (KAGRA) is
taking shape as workers construct twin sets of 3-km arms in new tunnels. Set
to start operation in 2018, KAGRA will use cryogenically cooled mirrors of
sapphire to deliver LIGO-like sensitivity.

\subsection*{References}

Many of the ideas expressed in this article have been discussed in the books
of Kennefick \cite{Ken07} and Collins \cite{Col04}.


\begin{thebibliography}{9}
\bibitem{Abb1} B.P. Abbott et al. Phys. Rev. Lett. \textbf{116}, 061102
(2016). \textit{Observation of Gravitational Waves from a Binary Black Hole
Merger}

\bibitem{Abb2} B.P. Abbott et al. Phys. Rev. Lett. \textbf{116}, 241103
(2016). \textit{GW151226: Observation of Gravitational Waves from a
22-Solar-Mass Binary Black Hole Coalescence}

\bibitem{Ken07} Daniel Kennefick, \textit{Traveling at the Speed of Thought:
Einstein and the Quest for Gravitational Waves} (Princeton University Press,
Princeton, 2007)

\bibitem{Col04} Harry Collins, \textit{Gravity's Shadow. The Search for
Gravitational Waves}. (University of Chicago Press, Chicago, 2004)
\end{thebibliography}
\end{document}